\documentclass[preprint2]{aastex}
\pdfoutput=1
\usepackage{epsfig}
\usepackage{amsmath}
\usepackage{hyperref}

\begin{document}

\title{Cosmography and Data Visualization}

\author{Daniel Pomar\`ede}
\affil{Institut de Recherche sur les Lois Fondamentales de l'Univers} 
\affil{CEA, Universit\'e Paris-Saclay, 91191 Gif-sur-Yvette, France}
\author{H\'el\`ene M. Courtois}
\affil{Universit\'e Claude Bernard Lyon I/CNRS/IN2P3, Institut de Physique Nucl\'eaire, Lyon, France}
\author{Yehuda Hoffman}
\affil{Racah Institute of Physics, Hebrew University, Jerusalem 91904, Israel}
\author{R. Brent Tully,}
\affil{Institute for Astronomy, University of Hawaii, 2680 Woodlawn Drive, Honolulu, HI 96822, USA}

\begin{abstract}

\smallskip
Cosmography, the study and making of maps of the universe or cosmos, is a field where visual
representation benefits from modern three-dimensional visualization techniques and media.
At the extragalactic distance scales, visualization is contributing in understanding the complex
structure of the local universe, in terms of spatial distribution and flows of galaxies and dark matter.
In this paper, we report advances in the field of extragalactic cosmography obtained using the SDvision
visualization software in the context of the Cosmicflows Project. Here, multiple visualization techniques
are applied to a variety of data products: catalogs of galaxy positions and galaxy peculiar velocities, 
reconstructed velocity field, density field, gravitational potential field, velocity shear tensor viewed
in terms of its eigenvalues and eigenvectors, envelope surfaces enclosing basins of attraction. 
These visualizations, implemented as high-resolution images, videos, and interactive viewers, have contributed
to a number of studies: the cosmography of the local part of the universe, the nature of the Great Attractor, the
discovery of the boundaries of our home supercluster of galaxies Laniakea, the mapping of the cosmic web,
the study of attractors and repellers.

\bigskip
\end{abstract}

\section{Introduction}

Throughout the ages, astronomers have strived to materialize their discoveries and understanding of the cosmos
by the means of visualizations. The oldest known depiction of celestial objects, the Nebra sky disc, dates back
from the Bronze age, 3600 years ago (Benson 2014). While most of the astronomical representations are
projections to two dimensional sketches and images, the introduction of the third dimension in
depictive apparatuses has been sought by astronomers as an essential mean to promote understanding.
Such objects as the armillary spheres, dating back from the Hellenistic world, or the modern era orreries
used to mechanically model the solar system, played an important role in the history of astronomy. Today,
computer-based interactive three-dimensional visualization techniques have become a fruitful research
tool. Here, we present the impact of visualization on cosmography in the context of the Cosmicflows Project.

The discipline of cosmography is both very old and very young. Since ancient times, societies have
tried to understand their place in the cosmos and have created representations to express their ideas.
Today we have learned that we live in an expanding universe, 13.7 billion years old, filled with
mysterious dark matter and dark energy, with galaxies in lattices of the cosmic web. However our
representations of the structure of the universe, cosmography, are remarkably primitive. We only
have good knowledge of a tiny fraction of the potentially visible universe.
The quest of the present research is to substantially improve our understanding of our local
neighborhood of the universe. Inevitably, what we know degrades with distance. However, nearby
our emerging picture is rich in detail and, farther away, the outer boundaries where our maps blur
into the background confusion are being pushed back dramatically.

The Cosmicflows Project aims at the reconstruction and mapping of structure of the Local Universe using peculiar
velocities of galaxies as tracers of the source density field; it has released three catalogs, 
Cosmicflows-1 with 1791 galaxies (Tully et al. 2008),  Cosmicflows-2 with 8161 galaxies (Tully et al. 2013),
Cosmicflows-3 with 17669 galaxies (Tully et al. 2016). This project has its foundations built on earlier works;
The Nearby Galaxies Atlas (Tully \& Fisher 1987) is a first cartography of the three dimensonal structure of the
Local Universe within redshifts of 3000 km/s, in association with the data documented in the Nearby Galaxies Catalog
(Tully 1988). This Atlas highlights the location of our galaxy on the periphery of the Local Void.
The distribution of galaxies is dominated by the presence of the Virgo cluster.
Then, using the union of Abell and Abell-Corwin-Olowin catalogs of rich clusters, a cartography
of structures extending to 30,000 km/s was published by Tully et al. (1992). Maps of the distribution of superclusters
are provided in the plane of the Supergalactic Equator that host the Great Attractor, the Great Wall,
the Shapley Concentration, the Pisces-Cetus Supercluster (Figure 4 in Tully et al. 1992). Maps of orthogonal slices
display the structure at higher SGZ altitudes with the Hercules, Corona-Borealis and Aquarius-Capricornus superclusters,
and at lower SGZ altitudes with the Lepus, Horologium-Reticulum and Sextans superclusters (Figures 3 and 5).
This early cartography was severely limited by the Zone of Avoidance associated with the obscuration caused by the
dust and stars of our own galactic disk. This Zone of Avoidance creates a three-dimensional wedge inside which direct
observations are not possible. It is a strength of the Cosmicflows program that it provides a reconstruction of the structures
lying within the Zone of Avoidance using the gravitational influence they exert on galaxies lying nearby.
Three-dimensional density contour maps within 8000 km/s were presented by Hudson (1993), see Figure 10 therein;
these show from two complementary viewpoints how filaments and high-density blobs such as Virgo-Centaurus-Hydra,
Perseus-Pisces, Pavo, Coma are organized. Using the 2M++ galaxy redshift compilation based on 2MRS, 6dFGRS-DR3 and
SDSS-DR7, Lavaux and Hudson (2011) presented maps of the density field within 15,000 km/s.
Using the Cosmicflows-1 Catalog, Courtois et al (2012) presented maps of both density and
velocity fields within 3000 km/s with reconstruction of the Local Void and the Great Attractor.

\section{The Cosmicflows program}

The objective of the Cosmicflows program is to map the distribution of matter and flows in the local
universe, understand the motion of our galaxy with respect to the CMB, and provide new insights in
cosmology by providing near-field measurement of basic parameters such as the Hubble Constant.
The fundamental ingredients of this program are the measurements of peculiar velocities of galaxies, that is their
deviation with respect to the Hubble expansion.
These peculiar velocities of galaxies serve as sensors of the source gravitational field, including dark matter. 
The Cosmicflows program has three critical constituents: observations, theoretical modeling, and
visualizations. Each of these is state-of-the art. 

On the observational side, the measurement of
accurate distances to galaxies permits the separation of deviant motions from the cosmic expansion (Tully et al. 2008, Tully et al. 2013, Tully et al. 2016).
Our program has accumulated, by far, the largest and most coherent assembly of galaxy distances.

The translation of measured radial velocities with substantial errors into a three-dimensional map
of galaxy motions is accomplished with a Wiener Filter technique averaging multiple realizations
constrained in a Bayesian analysis by the data and an assumed power spectrum of initial density
fluctuations (Zaroubi et al. 1999, Hoffman 2009, Courtois et al. 2012, Doumler et al. 2013). The analysis gives, in addition to the 3-D velocity field, the density and potential
fields, and the velocity field can be separated into local and tidal components. The shear of the
velocity field at each position defines eigenvectors and eigenvalues of the V-web, descriptors of
whether the location is in a knot, filament, sheet, or void (Hoffman et al. 2012). The analysis leads to the identification of
all the important basins of gravitational attraction and repulsion.

It is the role of the visualizations
to clarify the interplay between these various components in quite complicated circumstances. 
Cartography has played a seminal role in the development of our current understanding. With new
data sets, already arriving and projected, the visualization challenges are multiplying.

\section{The SDvision visualization software}

The SDvision (Saclay Data Visualization) software is deployed in the IDL ``Interactive Data Language" platform (Pomar\`ede et al. 2008). 
Originally developed in the context of the COAST ``Computational Astrophysics" Project for the visualization of
astrophysical simulations (Audit et al. 2006), it was realized that this software could also be used to visualize
cosmographic data (Pomar\`ede \& Pierre 2011, Pomar\`ede et al. 2013).
IDL was chosen for its widespread and long-term use in the Astronomy community, the professional support
and development plan offered by its owner company, the extensive astronomical libraries it offers, and for the 
high-performance visualization techniques it provides: IDL gives access to hardware-accelerated rendering techniques through
its ``Object Graphics" interfaces to OpenGL, as well as multi-threaded usage of multiple-core processors.
The graphics objects can be coupled with GLSL (OpenGL Shading Language) shader algorithms to perform both scientific
computation and visualization by graphics card. 
The SDvision visualization software consists today in 100,000 lines of code addressing the issues of strategic importance
in the field of cosmography: visualization of scalar fields, vector fields and clouds of points.

\begin{figure}[htbp]
\begin{center}
\includegraphics[scale=.435]{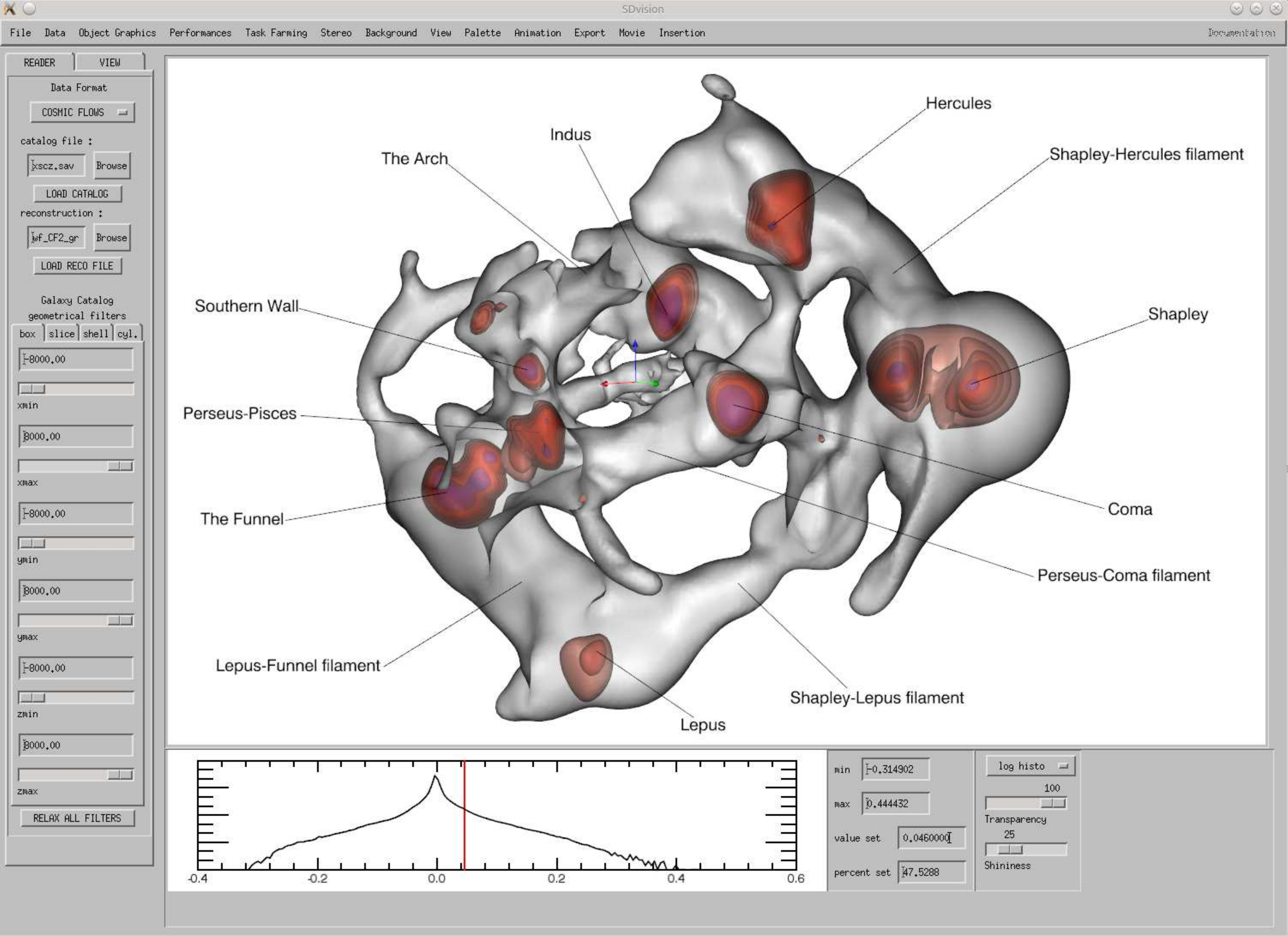}
\caption{The SDvision widget engaged in the visualization of multiple isosurfaces of the cosmic V-web.
The velocity web cartographied here is extracted from the analysis of the eigenvalues $\lambda_{i=1,2,3}$ of the velocity shear tensor. 
Five surfaces of constant values of the $\lambda_3$ eigenvalue, shown here in several nuances of red, trace the knots
of the web. A surface of constant value of the $\lambda_2$ eigenvalue, shown in gray, traces the filaments. The view is augmented with annotations and polylines indicating the most
salient cosmographic structures. These annotations and polylines are added to the transformable model like the other graphics objects.
A three-dimensional signpost made of three colored arrows (red, green, blue) is anchored on the origin of the Supergalactic coordinate system,
that is the location of the Milky Way,
with each arrow associated to its three cardinal axes (SGX, SGY, SGZ). 
}
\label{SDvision_widget_cosmicweb}
\end{center}
\end{figure}

The SDvision widget interface provides a multitude of tools to generate visual objects and act on their properties.
Built around a main view window where rotations, translations and scaling can be obtained by mouse click-and-drag actions,
the widget provides facilities relevant to cosmography such as sliders allowing on-the-fly geometrical cuts
on the galaxy catalog on display. The widget interface can be seen in Figure~\ref{SDvision_widget_cosmicweb} in the context of the visualization of scalar fields.
Here, the user is presented with a histogram. The user can click interactively on this histogram, an action that will result in the computation and updated visualization of the corresponding isosurface.
This interactive facility is most useful for the exploration of complex scalar fields, especially when coupled to the 3D navigation possibilities. 
The widget interface can also be seen in
Figure~\ref{SDvision_widget_xscz} in the context of the visualization of a catalog of galaxies.
A geometrical filtering is obtained by action on the sliders labelled xmin, xmax, ymin, ymax, zmin, zmax.
Among the $\sim 315,000$ galaxies of the catalog, $\sim 33,000$ are selected after applying these cuts.

Scalar fields, such as density or temperature fields, are visualized by the means of three
complementary techniques: 1) Ray-casting volume rendering. The ray-casting is a CPU-intensive
technique that propagates a ray through the volume under scrutiny and builds-up on contributions from
crossed cells, thus requiring a complete new computation at every change in the viewpoint. In this technique,
several composite functions are available to measure the value of a pixel
on the viewing plane by analyzing the voxels falling along the corresponding ray: the most basic
function is the Maximum Intensity Projection where the color of each pixel on the viewing plane is
determined by the voxel with the highest opacity value along the corresponding ray, the color of the
voxel being obtained against some lookup tables providing the three R, G, B colors and opacity functions. 
A more sophisticated compositing function is the Alpha blending, where a recursive
equation assumes that the color tables have been pre-multiplied by the opacity table to obtain semi-transparent
volumes. This technique is preferred where multiple layers of structures can hide each other. Finally the
ray-casting algorithm can be engaged to produce RGBA renderings where red, green, blue, and alpha
channels are associated to four different physical fields. The ray-casting algorithm is multithreaded,
exploiting all available computing cores on shared-memory
computing systems. 2) Isosurfaces reconstruction is used to display surfaces of constant value taken by
the scalar field. The reconstruction is performed by IDL's SHADE\_VOLUME procedure, that is similar to the Marching-cube algorithm.
The resulting surface is visualized as a Gouraud-shaded polygon. The interactive visualization of this polygon
benefits from the hardware acceleration by the graphics card. Figure~\ref{SDvision_widget_cosmicweb}
shows an example of surfaces reconstructed with this technique.
3) Slicing is used to map a texture on a simple slice of the volume under scrutiny, at any position and orientation.

Vector fields, such as velocity fields or magnetic fields, are visualized by two techniques: 1)
Streamlines reconstruction using IDL's PARTICLE\_TRACE procedure. This procedure traces the path of a massless particle through a vector field,
given a set of starting points (or seed points). Particles are tracked by treating the vector field as a velocity field and integrating.
Each particle is tracked from the seed point until the path leaves the input volume or a maximum number of iterations is reached.
The vertices generated along the paths are returned packed into a single array along with a polyline connectivity array used to feed a polyline
object. The collection of seeds can obey some predefined distribution (2D or 3D uniform grid,
spherical grid) or seeds can be selected interactively by clicking on any graphics object on display,
a most useful feature to study the structure of flows.
2) Hedgehog display where three-dimensional arrows are anchored on the same collection of points or seeds as in the streamlines procedure.

Point Clouds can be used to materialize the positions of galaxies obtained from catalogs. They are visualized using two techniques: 1) using
markers of definite sizes (polylines or polygons). This is useful to get varying apparent sizes of the
markers versus perspective and zooming. 2) Using sprites. In this technique a sprite shader forces the
adjacent pixels of the projected position of the points to participate in the rendering, resulting in
markers of fixed apparent sizes, independent of any perspective or zooming. This hardware-
accelerated technique is extremely fast and efficient for the rendering of larger numbers of cloud points.

The SDvision software can be controlled either interactively or through scripts describing a
sequence of commands. This latter option is used mainly to produce videos exposing the evolution in
a simulation or to explore a volume following predefined selected routes. It has facilities to
produce Stereo3D outputs and 8-cam outputs for autostereoscopic screens. The favored data structure
for scalar and vector fields is the uniform grid. More sophisticated data structures such as AMR
(Adaptive Mesh Refinement) are handled by projection onto a uniform grid of adequate resolution.
The algorithms implemented in SDvision favor the use of shared-memory architectures equipped
by multiple-core processors (to benefit from the ray-casting rendering
technique) and massive available RAM (to manage large datasets, in particular large 3D datagrids)
and with high-range graphics units (for the acceleration of polygon rendering and use of GLSL Shaders).

\section{Cosmography use cases}

In this section we illustrate the cosmographic potential of our software in four use cases: 1) the visualization
of galaxy catalogs, 2) the mapping of cosmic flows, 3) the visualization of three-dimensional basins of attraction, and 4) the
cartography of the cosmic web. 

\subsection{Visualization of catalogs of galaxies}

\begin{figure}[htbp]
\begin{center}
\includegraphics[scale=.435]{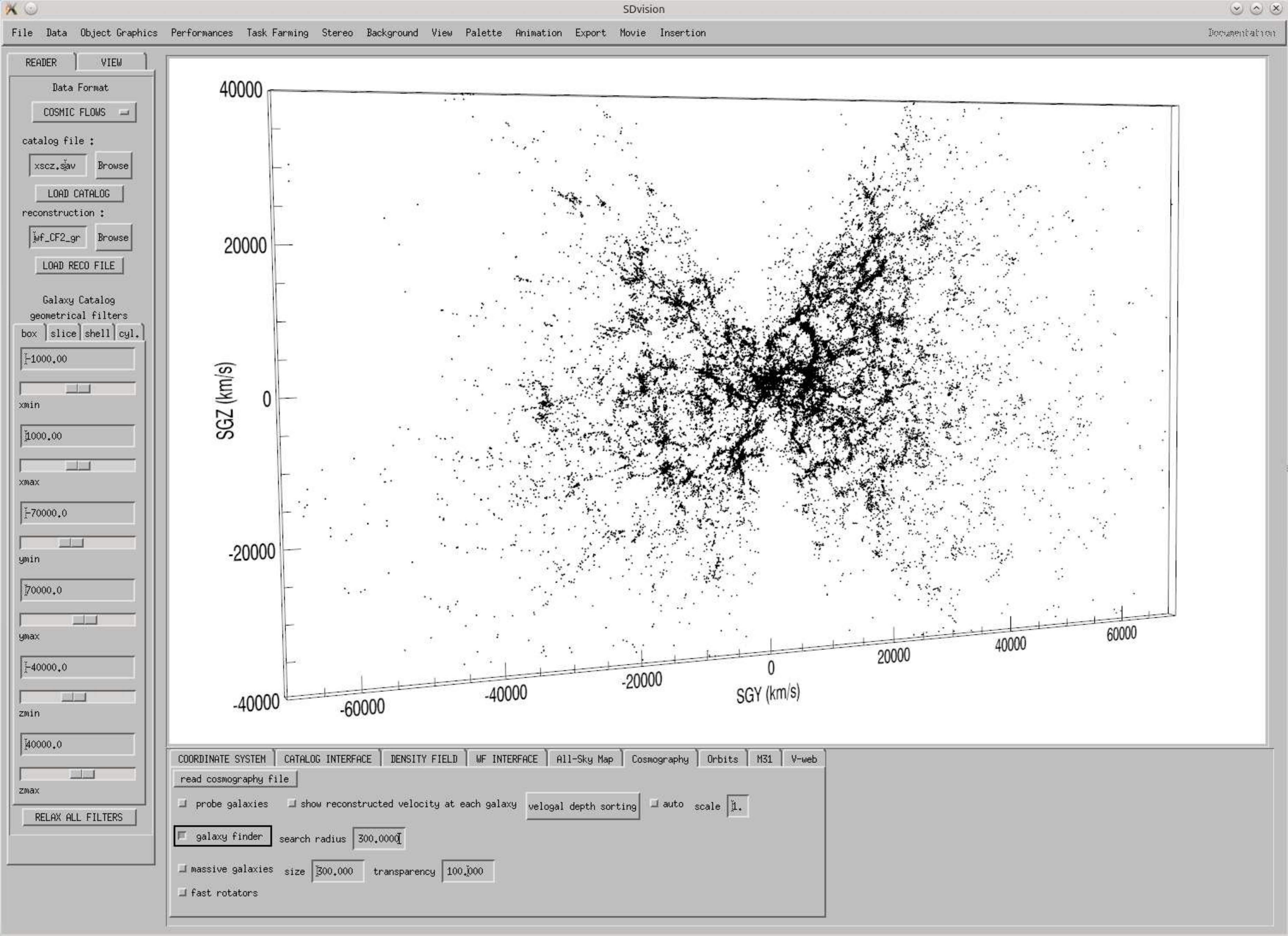}
\caption{The SDvision widget engaged in the visualization of a thin slice of the XSCz catalog of redshifts.
Our galaxy, the Milky Way, resides at the origin. The positions of galaxies are given in the Supergalactic coordinates system in units of redshift.
The butterfly like distribution is typical of the redshift surveys viewed in the SGY-SGZ and SGX-SGY projections, where the
obscuration by our galactic plane creates a three-dimensional wedge in the data sample where no measurements are available,
the so-called ``Zone of Avoidance".}
\label{SDvision_widget_xscz}
\end{center}
\end{figure}

Catalogs of galaxies may be produced that contain any number of galaxies, up to several millions. 
The visualization of their distribution is a key ingredient in cosmography. An example of running
a visualization of the XSCz catalog of redshifts (Jarrett et al. 2000) is displayed in Figure~\ref{SDvision_widget_xscz}.
In this example, the galaxies are mapped in the Supergalactic Coordinate System in units of redshift (km/s).
A thin slice -1000 km/s $<$ SGX $<$ +1000 km/s is selected.
The markers used to materialize the positions of the galaxies can either be spheres or sprites.
The rendering of many spheres is time-consuming, however it has the advantage of giving a physical size
to the galaxy markers; individual galaxies, or groups, or clusters can be approched during the exploration of the data,
with their individual object apparent size increasing during zoom-ins. On the other hand, sprites are rendered in real-time, but
they occupy a fixed number of pixels whatever the position of the eye; their apparent size does not change when zooming
in or out. In terms of cosmography, the SDvision interface provides useful tools: for instance the possibility to select
rectangular slices (using interactive sliders and fields visible in Figure~\ref{SDvision_widget_xscz} on the left-hand side
of the widget). Slices can be also oriented at any angle as a function of supergalactic longitute and latitude.
Spherical shells and cylindrical cuts can also be performed. The reference coordinate system can be displayed.
A cosmography-dedicated interface is proposed through which several tools are available: for example a galaxy finder, upon
activation, will find all the galaxies within a given search radius from any clicked point, and print their IDs, names
and positions. The visualization proposed in Figure~\ref{SDvision_widget_xscz} is typical of the maps obtained using redshift
surveys: it reveals a web of filaments connecting clusters of galaxies and separating empty voids. This is the cosmic web.
The first such maps were published 30 years ago (de Lapparent et al. 1986).
The 3D visualization of galaxy catalogs allows the exploration and the mapping of the structures in galaxy distribution 
(see e.g. Courtois et al. 2013). It is also useful to perform comparisons with reconstructed products that have complex 
three-dimensional architectures, such as the density field, basins of attraction, or the Cosmic V-web (see sections below).

In terms of visualization of galaxy catalogs, the Cosmicflows Project brings in an additional matter: the need to visualize
a radial velocity measurement tied to each galaxy position. Such ``peculiar velocities" are materialized as 3D vector arrows,
as exemplified in Figure~\ref{SDvision_vpec}. These vectors are the primary data input to the Wiener Filter algorithm that
reconstructs the fully three dimensional velocity field.

\subsection{Mapping cosmic flows}

\begin{figure}[htbp]
\begin{center}
\includegraphics[scale=.45]{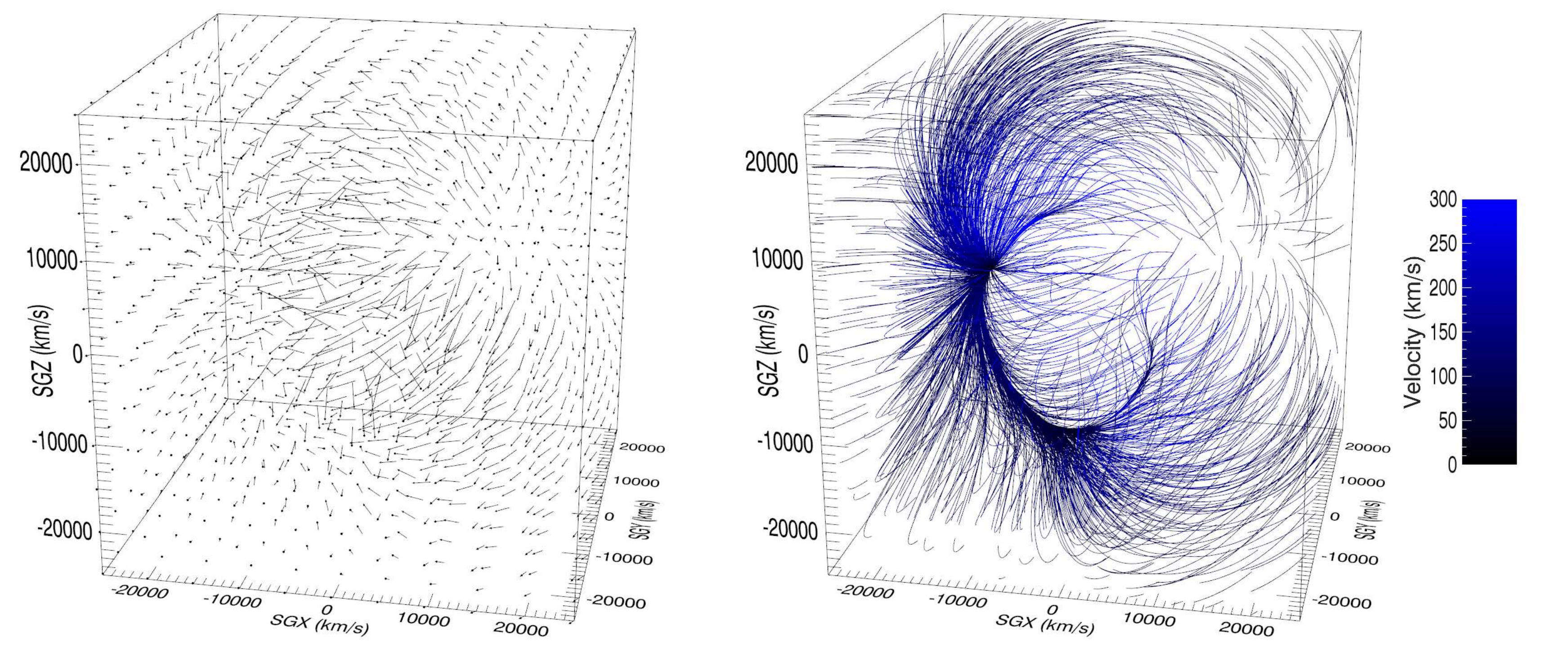}
\caption{Comparison of the visualization of a velocity field using arrows (left) and streamlines (right).
}
\label{SDvision_vectors_vs_streamlines}
\end{center}
\end{figure}

Cosmic flows are revealed in the visualization of the velocity vector field. Our favored means of visualizing a three-dimensional velocity
vector field is to use streamlines. Streamlines are polylines accounting for the reconstruction of a path from an initial ``seed" position
and proceeding by steps of constant size along the direction dictated by the local velocity arrow. The SDvision widget interface allows
interactive exploration of various step sizes (for a given line length, shorter step sizes require more computing time) and various total
streamline lengths. Short streamlines can be used to explore the structures in local computations of the flow field, while long streamlines
are useful to materialize the large scale structure manifested in the computations of the full flow. The visualization of streamlines offers
certain benefits over the standard vector-based representation (that is also implemented in SDvision).
The streamlines depict clearly and robustly the geometry of the flow.
The flow field (presented in Figure~\ref{SDvision_vectors_vs_streamlines}) is characterized by a convergence and by a divergent point,
that we call an attractor and a repeller. It further shows the filamentary nature of the flow.
The demonstration is much more difficult to realize from the arrows visualization. It is the continuity of the streamlines, and the
clear representation of the sources and sinks of the streamlines that make the difference.
In addition, the coloring of the streamlines provides a very clear visualization of the amplitude of the local velocity field. 

\begin{figure}[htbp]
\begin{center}
\includegraphics[scale=.185]{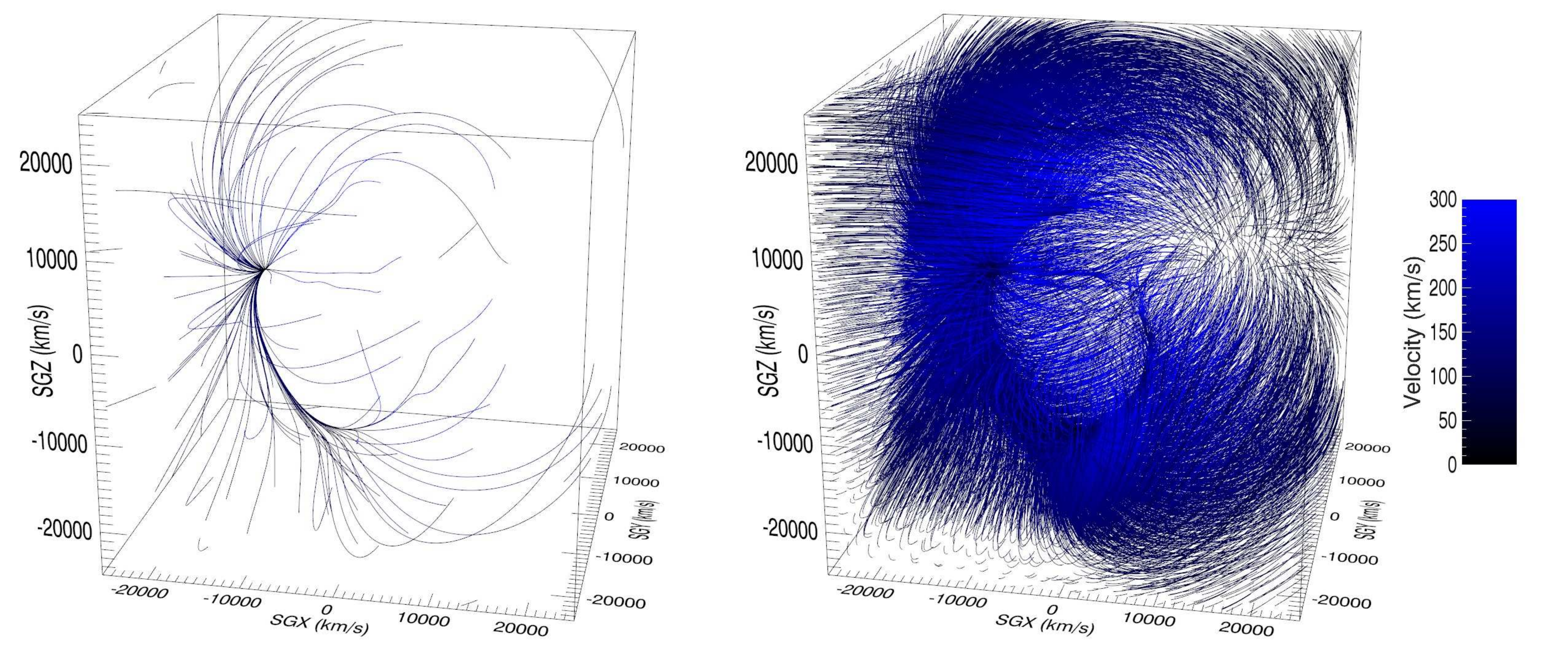}
\caption{Comparison of the visualization of a velocity field using streamlines seeded on 
uniform Cartesian grids of $5^3$ seeds (left) and $21^3$ seeds (right).
}
\label{SDvision_streamlines_seeding}
\end{center}
\end{figure}

\begin{figure}[htbp]
\begin{center}
\includegraphics[scale=.61]{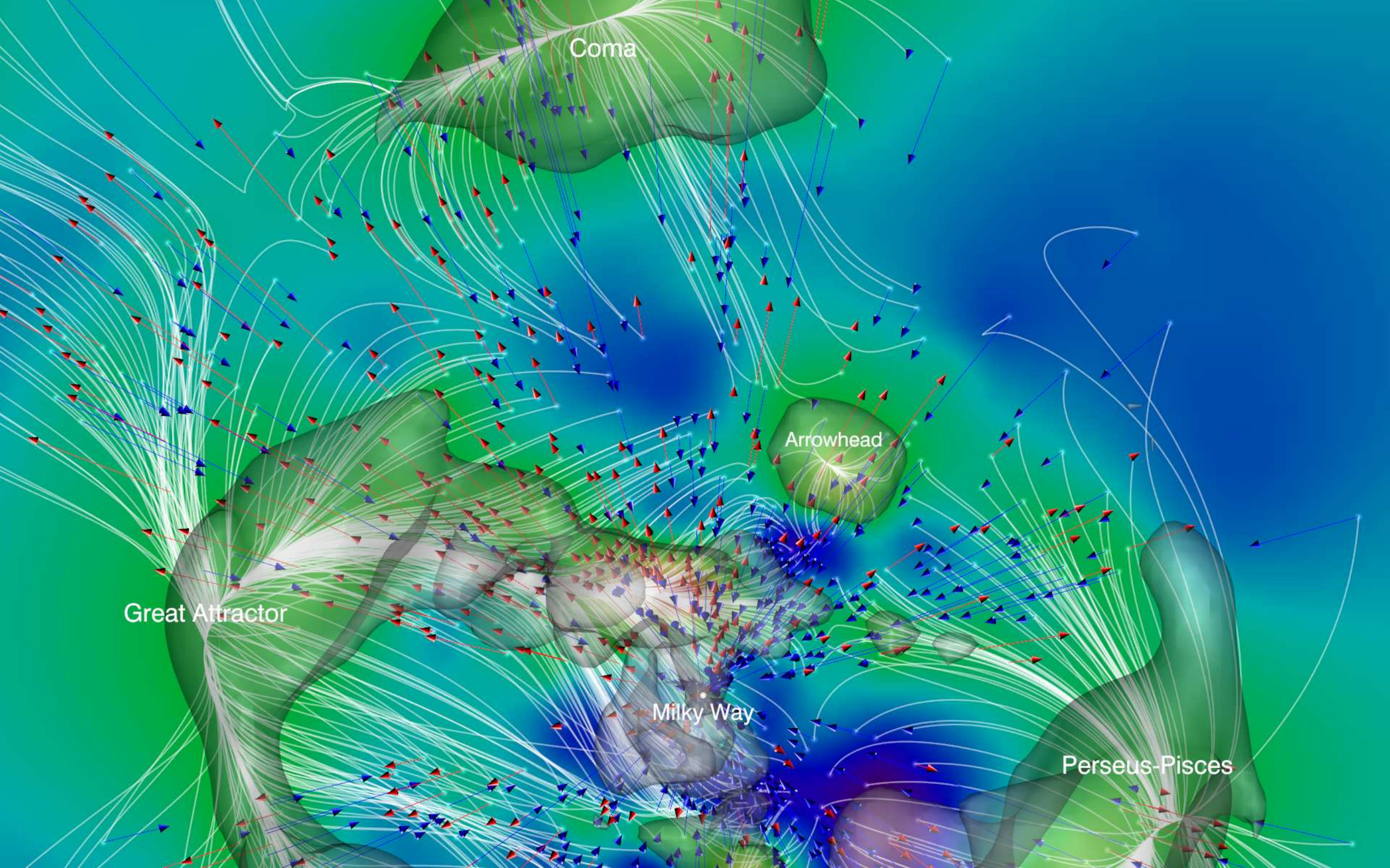}
\caption{Visualization of peculiar velocities and Wiener Filter products. The galaxies of the Cosmicflows-2 Catalog selected in a thin slice 
-500 km/s $<$ SGZ $<$ +1500 km/s are shown as white speckles. Their peculiar velocities are shown as three-dimensional arrows
colored blue for inward motions and red for outward motions. The white polylines are streamlines of the velocity field reconstructed by
the Wiener Filter, seeded at each galaxy location. These streamlines are ``long" streamlines, in the sense that their reconstruction
is pushed far enough so that each of them reach an attractor. The grey Gouraud-shaded polygon is an isosurface of the reconstructed density field. This high-density surface reveal the presence of the most massive players of the local universe:
the Great Attractor, Perseus-Pisces, and Coma. A more modest high-density patch, the Arrowhead, is also identified. A thin slice of the density 
field is also rendered as a rainbow-color contour image, showing nuances in the available densities, deep blue corresponding for example to voids
in the cosmography. This whole visualization conveys much of the essence of the Cosmicflows Project: peculiar radial velocities are used as input to
a Wiener Filter algorithm that produces a 3D velocity field, used itself to derive the density field and to search for structures in the distribution
of the streamlines, the visualization allowing the exploration of all these ingredients all together. Here the whole process results in the identification
of the Arrowhead basin of attraction inside which the streamlines converge on an attractor.}
\label{SDvision_vpec}
\end{center}
\end{figure}

An important aspect of streamline visualization is the seeding. The baseline configuration is to distribute the seeds on a Cartesian
3D uniform grid. This is an efficient way to obtain a global map of the cosmic flows. Such visualizations reveal essential structures in the flow:
the convergence on attractors for instance, or the compression of the lines into filaments. 
In Figure~\ref{SDvision_vectors_vs_streamlines}, the seeds are located on the nodes of a $11^3$ uniform
Cartesian grid, providing a reasonably nuanced view of the global structure of the flow.
In Figure~\ref{SDvision_streamlines_seeding}, left, a lighter sample of $5^3$ seeds is used.
This seeding does a good job at finding an attractor but does not inform much on the rest of the volume.
In Figure~\ref{SDvision_streamlines_seeding}, right, a heavier sample made of $21^3$ seeds is used. Here the view is saturated with information. These examples illustrate how the seeding affects the visualization.

\begin{figure}[htbp]
\begin{center}
\includegraphics[scale=.615]{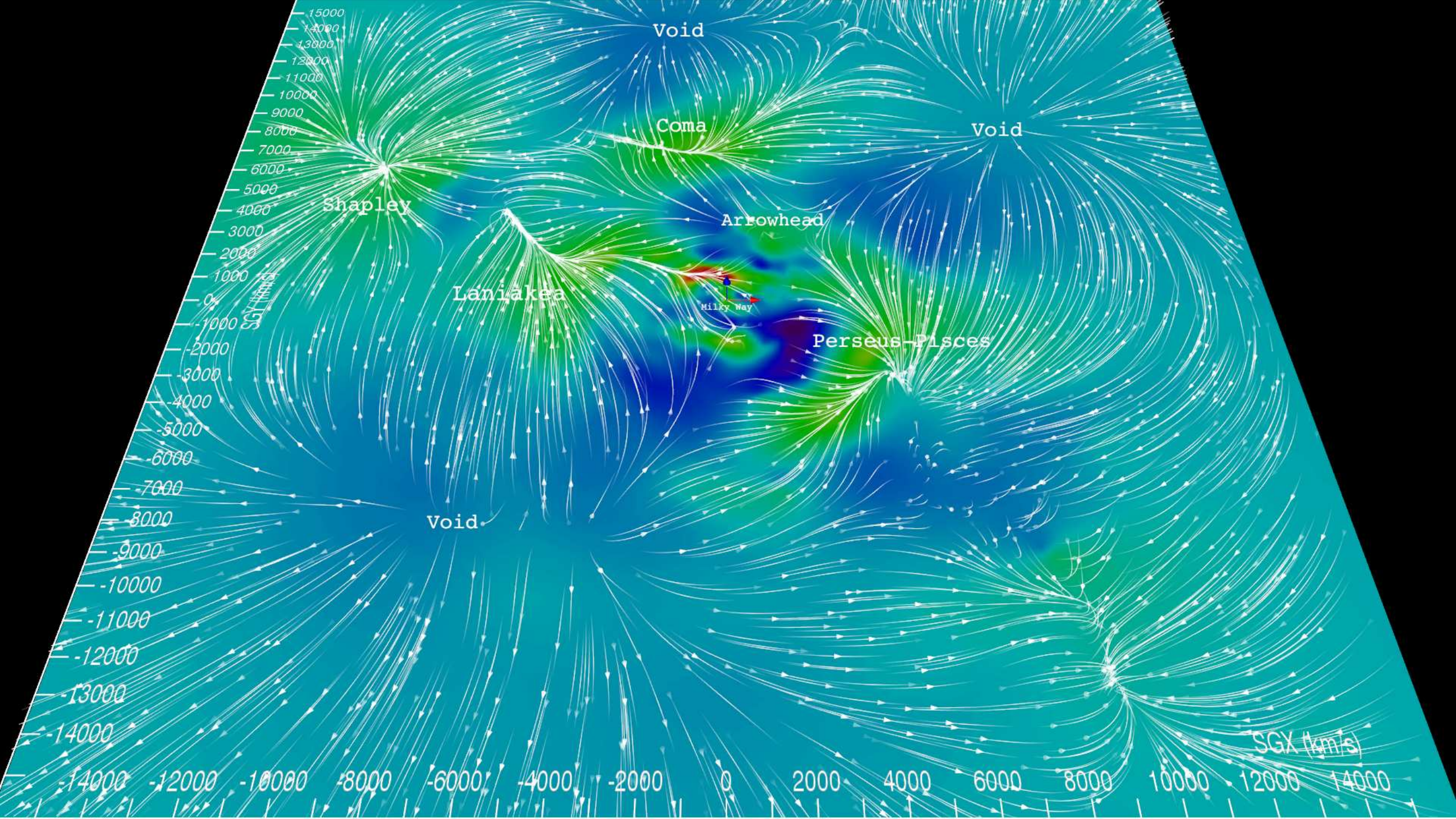}
\caption{Visualization of cosmic flows. This map of the Supergalactic equatorial plane (SGZ=0) is obtained using ``short" streamlines:
short in the sense that their reconstruction is halted at a given number of steps without pursuing the convergence onto an attractor.
Here, each single streamline benefits from it own local 3D velocity field, obtained by separation of the local (divergent)
component of the velocity field from the external (tidal) component. This combination of using a local field and short streamlines results in the
mapping of very local flows (at the cost of a single arbitrary parameter, the radius of the sphere of divergent/tidal separation).
This process reveals a structure of adjacent basins of attraction, whether fully reconstructed (Laniakea, Arrowhead) 
or partially reconstructed (Perseus-Pisces, Coma, Shapley) due to their location on the edges of the data sample. This map is also powerful
in illustrating the concept of evacuation of the flow from the voids. Also shown is a rainbow-color contour image of the density field that
provides additional context: the basins of attraction are found to be associated with higher-density environments. Frontiers between adjacent
basins of attractions are running through underdense regions.}
\label{SDvision_localflows}
\end{center}
\end{figure}

\begin{figure}[htbp]
\begin{center}
\includegraphics[scale=1.]{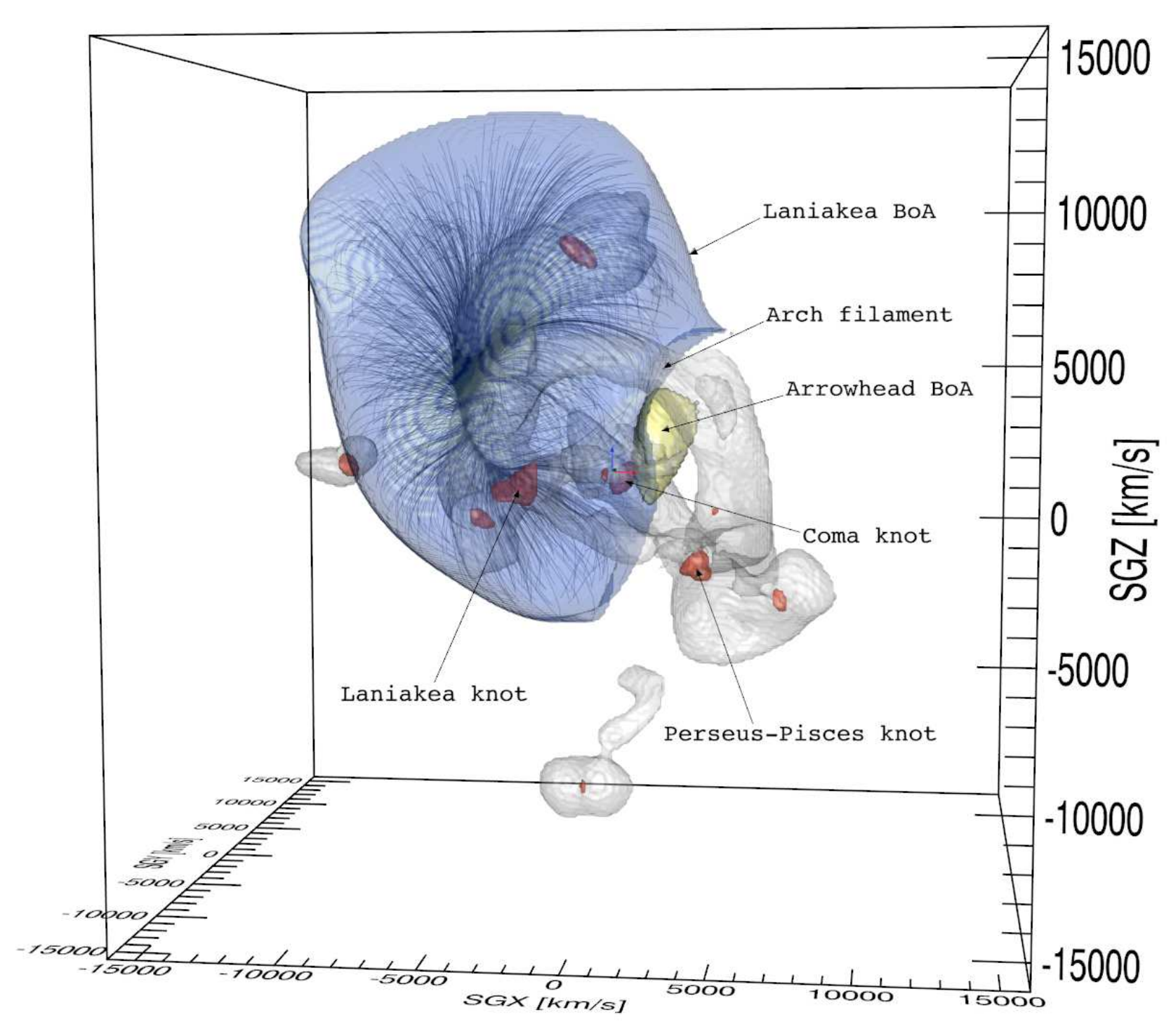}
\caption{Visualization of basins of attraction. The two surfaces enclosing the basins of attraction of Laniakea and Arrowhead are shown as
blue and yellow semi-transparent Gouraud-shaded polygons. Streamlines seeded randomly within the Laniakea envelope are shown as black
polylines. To provide more context, the V-web is also represented with a grey surface for the filaments and a red surface for the knots.
In terms of cosmography, it is particularly interesting to examine the Arch filament that connects the Perseus-Pisces knot to the Laniakea knot, crossing
the Laniakea border at a normal incidence.
}
\label{SDvision_BoAs}
\end{center}
\end{figure}

Since the seeding along three directions can be
the source of confusion between lines located at various depths, it is interesting to restrict the seeding to planes. The SDvision widget allows one
to interactively scan through the volume by seeding the streamlines on planes normal to each cardinal direction. The range and number of seeds
can be tuned interactively to optimize the information, keeping in mind that too many streamlines can result in confusion, while too few can result
in missing essential information. Another way to seed the streamlines is to use an ancillary source. For instance, the position of galaxies
from any catalog can be used as seeds. This is what is achieved in Figure~\ref{SDvision_vpec}.
This is especially interesting in the context of
the Wiener Filter analysis, as we can visualize together the input peculiar radial velocity of the algorithm, and its output fully 3D reconstructed
peculiar velocity. Another type of seeding is to take at random uniform distribution, for instance in a 3D volume, or confined to a 2D plane.
This latter option is illustrated in Figure~\ref{SDvision_localflows} with short streamlines, providing a map
of local cosmic flows in the plane of the Supergalactic Equator. Finally, it is also possible to combine randomness and ancillary constraints,
by e.g. distributing seeds randomly inside or outside an isosurface of any given scalar field, or any other volume defined in any mathematical way.
An example of such seeding is presented in Figure~\ref{SDvision_BoAs} where seeds are distributed randomly and uniformly within the surface 
of the Laniakea basin of attraction (see following section).

\subsection{Basins of attraction and the definition of superclusters}

A basin of attraction is a volume inside which the flow exhibits convergence onto a unique point.
This notion emerged when this convergence was first observed in the Wiener Filter reconstruction of the
Cosmicflows-2 Catalog (Tully et al. 2013). It was proposed to use this notion as a definition for superclusters of galaxies, and
the application of this idea resulted in the identification of the Laniakea supercluster, our
home supercluster (Tully et al. 2014). The visualization of basins of attraction is illustrated in Figure~\ref{SDvision_BoAs},
where the Laniakea basin of attraction is displayed together with the Arrowhead basin of attraction
(Pomar\`ede et al 2015) as semi-transparent
polygons. The concurrent visualization of streamlines of the velocity field seeded randomly within the
Laniakea basin of attraction demonstrates the convergence of the flow onto a single attractor.
The structure of the Laniakea basin of attraction
in the plane of the Supergalactic Equator can also be appreciated in Figure~\ref{SDvision_localflows} where
the tiling of the universe in adjacent basins of attraction can be inferred.

\subsection{The Cosmic V-Web}

The analysis of the velocity field in terms of expansion and compression of the flow provides the signatures
of a web-like structure called the Cosmic Velocity Web, or V-Web (Hoffman et al. 2012). At each point
in space, four possible situations are identified:
\begin{itemize}
\item a compression of the flow observed in three orthogonal directions is a signature
of a knot. Such a knot is acting as an attractor where the flow is converging. 
\item a compression of the flow observed in two orthogonal directions and an expansion
observed in the third direction is a signature of a filament. The filament collects the flow
from its surroundings and direct it along a single direction.
\item an expansion in two directions and compression in the third direction is the signature
for a sheet. The flow runs freely along two directions and experience compression in
the third.
\item an expansion in all three orthogonal directions is a signature of a void, a region that can be
considered as a repeller from which an expulsion is ongoing. 
\end{itemize}

\noindent
These signatures are expressed mathematically by the velocity shear tensor.
The velocity shear tensor is characterized in terms of its three eigenvalues ordered as follows:
$\lambda_1 > \lambda_2 > \lambda_3$. 

\noindent
The mathematical prescriptions for the knots, filaments, sheets, and voids are the following:

$\lambda_3 > {\lambda_3}_{threshold} > 0$ is signature of a knot.

$\lambda_2 > {\lambda_2}_{threshold} > 0$ is signature of a filament. 

$\lambda_2 < {\lambda_2}_{threshold} < 0$ is signature of a sheet. 

$\lambda_1 < {\lambda_1}_{threshold} < 0$ is signature of a void. 

Figure~\ref{SDvision_widget_cosmicweb} presents a visualization of the V-Web in terms of
its knots and filaments, using multiple surfaces of constant values
of $\lambda_3$ and $\lambda_2$. 
The emerging structure is that of a three-dimensional web of knots connected by filaments.
This reconstruction includes uncharted territories where direct observations are not permitted,
such as the zone of avoidance.

\section{Video productions}

If an image can be worth a thousand words, a video is worth many thousand (Pomar\`ede et al 2015).
Through a series of rotations, translations, zooms, the viewer of the visualization video follows structures in three dimensions
and can grasp the relationships between features on different scales while retaining a sense of
orientation. Various graphics objects (polygons, polylines, clouds of points) materializing various
fields (density, velocity, V-web elements, gravitational potential, galaxies,...) and other entities (basins
of attraction)
can be represented simultaneously to study their relationships. In case of complex configurations of objects
where confusion can arise, the fading-in and out of certain objects in the course of the video can promote
deeper understanding.

Several visualization videos have been produced as part of the peer-reviewed publications listed in the sections below.
These videos are released on Vimeo and YouTube and can also be accessed through dedicated web pages hosted by CEA/IRFU.
The videos can either be viewed on internet browsers or downloaded in several resolutions fitted to most devices:
HD 1080p, HD 720p, SD, Mobile SD. In some cases, Stereo3D versions are available and comments are available as CC ``Closed
Captions" that can be also downloaded separately.

\subsection{Cosmography of the Local Universe}

This 17min35s video with comments is presented at:

\noindent
\href{http://irfu.cea.fr/cosmography}{http://irfu.cea.fr/cosmography}

\noindent
It explores the cosmography of the local universe by exploring the 22 maps presented in Courtois et al. 2013.
The reconstructions of the velocity field are based on the Cosmicflows-1 Catalog of peculiar velocities.
This video obtained 410,000 views so far and was downloaded 10,000 times.
  
\subsection{Laniakea Supercluster}

This 7 minutes-long video with comments is available at:

\noindent
\href{http://irfu.cea.fr/laniakea}{http://irfu.cea.fr/laniakea}

\noindent
It presents the identification and characterization of the Laniakea Supercluster of galaxies. It has been published as part of
Tully et al. 2014.
The reconstructions of the velocity field are based on the Cosmicflows-2 Catalog of peculiar velocities.
This video was viewed 170,000 times so far in its various forms.

\subsection{Laniakea Nature Video}

By using extracts of the two previous videos, the Nature Publishing Group has produced
the following 4min11sec video:

\noindent
\href{https://youtu.be/rENyyRwxpHo}{https://youtu.be/rENyyRwxpHo}

\noindent
This video has been released on YouTube's Nature Video Channel:

\noindent
\href{https://www.youtube.com/user/NatureVideoChannel}{www.youtube.com/user/NatureVideoChannel}

\noindent
As of september 2016, this channel has 150,000 subscribers and harbors 316 videos produced over the course
of the past eight years, for a grand total of 33.6 million views. The Laniakea Video obtained a record
of 500,000 views in a single day upon its release and became by far the most popular video with a total
of more than 4 millions views, 30,000 likes and 2,100 comments.
  
\subsection{The Arrowhead Mini-Supercluster}

This 4min18s video is available at:

\href{http://irfu.cea.fr/arrowhead}{http://irfu.cea.fr/arrowhead}

It explores the Arrowhead supercluster, a basin of attraction of limited size located at the contact region of the
Laniakea supercluster and its two giant neighbors Perseus-Pisces and Coma (Pomar\`ede et al 2015).
This video has been viewed 600 times so far.

\section{Interactive web-based visualizations}

We are exploring the use of Sketchfab, a web-based service that enables the upload and sharing of 3D models.
It is an efficient tool to share visualizations among collaborators. 
The Sketchfab interactive viewer is based on WebGL, and can be embedded like a video player in any html document.
This new technology thus offers a new way to visualize data interactively in scientific articles.

An example of such an interactive visualization reproducing the Figures 8 and 9 of Courtois et al. 2013 can
be viewed here:

\noindent
\href{https://skfb.ly/R9pN}{https://skfb.ly/R9pN}

This visualization shows the rendering of three isosurfaces of the reconstructed density field of the Local Universe. Navigation basics include rotations (left click and drag), zoom (scroll mouse wheel when available, or Ctrl+Left click and drag up and down),
pan (right click and drag). More sophisticated hints and controls are available by clicking on the question mark icon provided by the player.
Annotations are used to mark the most relevant structures. The navigation capabilities allow
an extensive exploration of the cosmography.

Technically, objects such as polygons, polylines, textures, can be uploaded in Sketchfab in a variety of different formats.
We use the Wavefront OBJ file format. The SDvision software has a facility to dump polygon objects such as isosurfaces
to OBJ files. More sophisticated writer codes are developed in IDL.

\begin{figure}[htbp]
\begin{center}
\includegraphics[scale=.272]{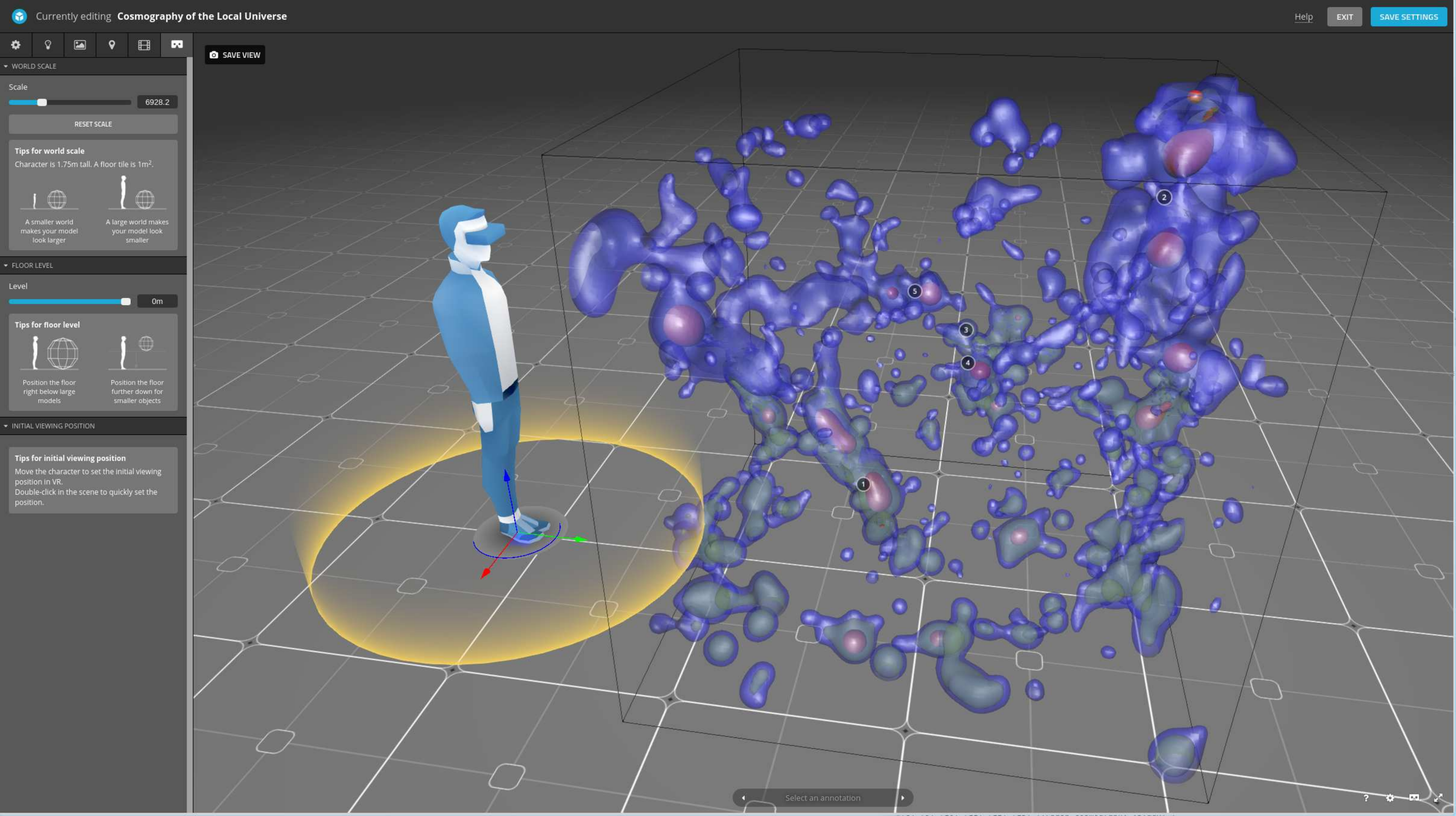}
\caption{The Sketchfab VR interface. The scene on display features isosurface objects reproducing the Figures 8
and 9 of Courtois et al. 2013. The corresponding interactive web-based visualization can be viewed here:
\href{https://skfb.ly/R9pN}{https://skfb.ly/R9pN}. This VR interface provides tools to modify scale, floor level,
and initial viewing location. Other 3D properties such as lighting, camera field of view, materials, animations, annotations can
also be fine-tuned using dedicated interfaces.
}
\label{SDvision_Sketchfab}
\end{center}
\end{figure}

The Sketchfab platform offers an advanced interface to fine-tune the 3D properties of the uploaded objects. It also
provides Virtual Reality (VR) in association with VR gears and VR helmets. The VR interface in action in the context 
of a cosmography visualization is shown in Figure~\ref{SDvision_Sketchfab}; here the interface provides a series
of interactive widgets to set the initial viewing position, the world scale, and the floor level.

\section{Conclusion}

Data visualization is a key ingredient in the field of extragalactic cosmography.
It promotes understanding of the three-dimensional structure of the cosmos and
can be part of the discovery process.
The development of our visualization software has accompanied the developments
in observations and theory. Together they contributed to advances in the field, as evidenced
in this paper.

Cosmography is an ancient science that responds to a deep desire within societies to answer the
philosophical question of our place in the Universe. It perpetuates the effort mankind has made in
mapping the cosmos. From a societal viewpoint, the releases of the previous generations of
cosmographical maps have demonstrated a profound enthusiasm in society as demonstrated by the
audiences reached by diverse media. The video ``Laniakea: Our home supercluster" produced by
the Nature Publishing Group is ranked as number one in popularity on the ``Nature Video" channel on YouTube with
over 4 million views. There have been an abundance of articles in news outlets, popular
science media and blogs, social networks and direct contacts with both the public and scientists.

\section{References}

\noindent
Audit, E., et al., 2006, ASPCS, 359, 9

\noindent
Benson, M. 2014, Cosmigraphics - Picturing space through time (Abrams, New York) 71

\noindent
Courtois, H., et al. 2012, ApJ, 744, 43

\noindent
Courtois, H., et al. 2013, AJ, 146, 69

\noindent
de Lapparent, V., Geller, M.J., \& Huchra, J.P. 1986, ApJ, 302L, 1 

\noindent
Doumler, T., et al., 2013, MNRAS, 430, 888

\noindent
Hoffman, Y., 2009, in Lecture Notes in Physics, Berlin Springer Verlag, Vol. 665, Data Analysis in Cosmology, ed. V.J. Mart\'inez, E. Saar, E. Mart\'inez-Gonz\'alez, \& M.-J. Pons-Border\'ia, 565-583

\noindent
Hoffman, Y., et al., 2012, MNRAS, 425, 2049


\noindent
Hudson, M.J., 1993, MNRAS, 265, 43

\noindent
Jarrett, T.H., et al., 2000, AJ, 119, 2498

\noindent
Lavaux, G., \& Hudson, M.J., 2011, MNRAS, 416, 2840

\noindent
Pomar\`ede, D., \& Pierre, M., 2011, IAUS, 277, 154

\noindent
Pomar\`ede, et al., 2008, ASPCS, 385, 327

\noindent
Pomar\`ede, D., Courtois, H., Tully, R.B., 2013, IAUS, 289, 323

\noindent
Pomar\`ede, D., et al. 2015, ApJ, 812, 17

\noindent
Tully, R.B., \& Fisher, J.R., Nearby Galaxies Atlas, Cambridge University Press, 1987

\noindent
Tully, R.B., Nearby Galaxies Catalog, Cambridge University Press, 1988

\noindent
Tully, R.B., et al. 1992, ApJ, 388, 9

\noindent
Tully, R.B., et al. 2008, ApJ, 676, 184

\noindent
Tully, R.B., et al. 2013, AJ, 146, 86

\noindent
Tully, R.B., et al. 2014, Nature, 513, 71

\noindent
Tully, R.B., et al. 2016, AJ, 152, 50

\noindent
Zaroubi, S., Hoffman, Y., Dekel, A., 1999, ApJ, 520, 413

\onecolumn

\twocolumn

\end{document}